\documentclass[
    ,final            
    ,numberedheadings 
  ]
  {aipproc}
\usepackage{graphicx}
\usepackage{amsmath}
\usepackage{amssymb}

\layoutstyle{6x9}

\begin{document}
\title{Non-exotic theory of 1/f noise as a trace of infralow-frequency fluctuations}

\author{A. Ya. Shul'man}{
  address={Institute of Radio Engineering and Electronics, Russian Academy of Sciences,
\\
Moscow 125009, Russia. E-mail: ash@cplire.ru} }

\begin{abstract}
This report is aimed at reviving the explanation of flicker-noise observations 
as the result of spectral measurement of very low-frequency but stationary 
narrow-band fluctuations named as infralow-frequency noise (ILF noise) [A. Ya. 
Shul'man. Sov. Tech. Phys. Lett. 7, 337 (1981), Sov. Phys. JETP 54, 420 
(1981)]. Such a kind of the spectrum analyzer output takes place when the 
ILF-noise correlation time is much longer than the analyzer reciprocal 
bandwidth. This result is valid for both analog and digital spectral 
measurements. The measured signal is proportional in this case to the mean 
square and not to the spectral density of the noise. The equilibrium 
temperature fluctuations and the defect-motion as mechanisms of $1/f$ noise in 
metal films are reconsidered from this point of view. It is shown that the 
ILF-noise approach allows to remove the main objection against temperature 
fluctuation model and difficulties within the defect-motion model widely 
discussed in literature. 
\end{abstract}

 \maketitle
\section{Introduction}
It was shown in \cite{AShTPL81}-\cite{AShJETP81} that
flicker-noise observations can be explained as the result of
spectral measurement of very low-frequency but stationary
narrow-band fluctuations named as infralow-frequency noise (ILF
noise). Such a kind of the spectrum analyzer output takes place
when the measured noise spectrum concentrates on frequencies far
lower than the operation range of the spectrum analyzer and the
ILF-noise correlation time is much longer than the analyzer
reciprocal bandwidth. This result is valid for both analog and
digital spectral measurements. The measured signal is proportional
in this case to the mean square rather than the spectral density
of the noise, and does not reflect in any way the dependence of
the latter on the frequency. Within the scope of this model the
universality of the $1/f$ noise follows from the universality of
inevitable fluctuations of intensive thermodynamic variables like
the temperature and a power-low falling of narrow-band filter
spectral distributions in the wings. It is of importance here the
more isolated is the sample under study from the environment the
more low-frequency are corresponding fluctuations. Thus, it is
impossible to reach the low-frequency saturation region of the
noise spectral density if the duration of the measurements is less
than response time of the sample - environment coupling.

The necessity to find not only slow fluctuation process but also
with $1/f$ -spectrum serves as the reason to call in question some
physical mechanisms that by their qualitative attributes and
quantitative estimations could be a cause of the excess current
noise. Example of this is the temperature model of $1/f$ noise in
metal films suggested by Voss and Clarke on the basis of
convincing experiments \cite{V-C PRB76p556}. Taken initially with
hopes and an enthusiasm \cite{ShK UFN77p131}-\cite{D-H RMP81p497}
this theory has later been rejected as an explanation of $1/f$
noise in metal films \cite {ShK UFN85p285},\cite{Weiss RMP88p537}.
In addition to the proof that the heat conductivity equation does
not lead under no circumstances to the $1/f$ -spectrum of
temperature fluctuations \cite{LVain JETP82p1841}, another weighty
argument against this model was the absence of spatial
correlations of the current fluctuations \cite{SDW
PRB81p7450},\cite{BWF PRB24p7454}.

On the contrary, there is no need in the scope of the ILF-theory
for exponentially-wide distribution of relaxation times of any
kind \cite {AShJETP81}. The interpretation of the spatial
correlation measurements is also to be reconsidered since the mean
square of the conterminous fluctuations is measured in the case of
ILF-noise rather than the spectral density of them. It will be
shown here that all abovementioned objections against temperature
fluctuations are withdrawn from the ILF-noise point of view.
\section{The noise spectroscopy and infralow-frequency fluctuations}
\subsection{Analogue measurements}
It was shown in \cite{AShTPL81}-\cite{AShJETP81} that the output $%
G^{out}(\omega _{0})$ of an analogue spectrum analyzer as a
function of the resonant frequency $\omega _{0}$\ can be described
in a good approximation by the convolution integral
\begin{equation}
G^{out}(\omega _{0})=\frac{1}{\Delta }\int\limits_{-\infty }^{\infty }\frac{%
d\omega }{2\pi }{\cal K}\left( \frac{\omega _{0}-\omega }{\Delta
}\right) G_{y}(\omega ),\,\Delta /\omega _{0}\ll 1,\,0<\left|
\omega _{0}\right| <\infty ,  \label{Eq-base}
\end{equation}
(the notation is slightly changed). Here $G_{y}(\omega )$ is the
spectral density of the stationary random input process $y(t)$
with $\left\langle
y^{2}\right\rangle <\infty $, ${\cal K}\left( \frac{\omega _{0}-\omega }{%
\Delta }\right) $ plays the role of the instrumental function of
the
spectrum analyzer normalized by the condition ${\cal K}\left( 0\right) =1$, $%
\Delta $ is the effective bandwidth of the analyzer defined by
\begin{equation}
\frac{1}{\Delta }\int\limits_{-\infty }^{\infty }\frac{d\omega }{2\pi }%
{\cal K}\left( \frac{\omega }{\Delta }\right) =1.  \label{Norm}
\end{equation}

An important point is that ${\cal K}(\omega )$ and $G_{y}(\omega
)$ are nonnegative and normalizable functions of the frequency and
can be approximated by $\delta $-function each at proper
interrelation between the effective bandwidths. If ${\cal
K}(\omega )$ is noticeably more narrow-band function then
\[
G^{out}(\omega _{0})=G_{y}(\omega _{0})
\]
and the result of measurement $G^{out}(\omega _{0})$ is the
common-adopted estimation of the true spectral density
$G_{y}(\omega )$. On the other hand if the investigated spectrum
has a narrower contour than the instrumental function, then the
result of the measurements will be the contour of the instrumental
function
\begin{equation}
G^{out}(\omega _{0})=\frac{1}{\Delta }{\cal K}\left( \frac{\omega
_{0}-\omega _{m}}{\Delta }\right) \left\langle y^{2}\right\rangle
, \label{ILF-analog}
\end{equation}
where $\omega _{m}$ is the frequency at which the maximum of
$G_{y}(\omega )$ takes place. Spectra are always measured with a
narrow-band filter, when the quality factor or the figure of merit
\begin{equation}
Q\equiv f_{0}/\Delta \gg 1,\,(f_{0}=\omega _{0}/2\pi )
\label{Q-defin}
\end{equation}
. It is now obvious from Eq. (\ref{ILF-analog}) that when the
spectral density of the noise is narrower than ${\cal K}(\omega
)$, and the noise frequency is low enough $\omega _{m}\ll \omega
_{0}$ [this is the definition of the term ''infralow-frequency''
noise], then in the usual $Q=const(\omega _{0})$ measurements the
output should have a dependence of the $1/f$ type:
\begin{equation}
G^{out}(\omega _{0})=\frac{Q}{f_{0}}{\cal K}\left( 2\pi Q\right)
\left\langle y^{2}\right\rangle .  \label{1/f-analog}
\end{equation}
On the other hand, if the measurements are carried at $\Delta
=const(\omega _{0})$, then the output signal has the frequency
dependence of the far wing of ${\cal K}(\omega )$, which likewise
has as a rule the form $1/f^{n}$. In either of these case the
measured values are proportional to the mean square $\left\langle
y^{2}\right\rangle $\ of the input random process.
\subsection{Digital spectral analysis}
In some respect the analysis of digital methods, from our point of
view, is even simpler, since the formula similar to Eq.
(\ref{Eq-base}) is already known. There is the exact expression of
the form
\begin{equation}
G^{out}(\omega _{0})=T_{0}\int\limits_{-\infty }^{\infty }\frac{d\omega }{%
2\pi }{\cal W}\left( T_{0}(\omega _{0}-\omega )\right)
G_{y}(\omega ),\,\,\,-\infty <\omega _{0}<\infty ,
\label{Dig-base}
\end{equation}
where $T_{0}$ is the time slice of the one realization of the
random process and$\ {\cal W}(\omega )$ is the spectral window.
Widely used windows can be written in the form
\begin{equation}
{\cal W}(\omega )\propto \left[ \frac{\sin \left( \omega T_{0}/n\right) }{%
\omega T_{0}/n}\right] ^{n}  \label{W-FT}
\end{equation}
where $n=2,4$ corresponds to the Bartlett and Parzen window,
respectively.

The account for
 finiteness of the time step $\tau _{0}$ and
application of discrete Fourier transformation lead to a little
bit changed expressions for spectral windows. For Bartlett window
one has \cite {AShJETP81}
\begin{equation}
{\cal W}_{d}(\omega )=\frac{\sin ^{2}\left( \omega N\tau _{0}/2\right) }{%
N^{2}\sin ^{2}\left( \omega \tau _{0}/2\right) }  \label{WB-discr}
\end{equation}
and $T_{0}=N\tau _{0}$, $N$ is the number of data points in the
measured realization. So long as the frequency $\omega $ is small
compared with the Nyquist frequency, defined by the condition
$\omega _{N}=2\pi f_{N}=\pi /\tau _{0}$ the continuous and
discrete Fourier transformations yield practically identical
expressions for the instrumental function (spectral window).
However, since ${\cal W}_{d}$ is the periodical function of
$\omega $, the formula (\ref{Dig-base}) can give rise to the
aliasing effect. To avoid this, the input process $G_{y}(\omega
)$\ should be filtered by low-pass filter with cut-off frequency
of order of $f_{N}$. With this
condition in mind, the comparison of Eqs.(\ref{Eq-base})-(\ref{Norm}) to Eqs.(%
\ref{Dig-base})-(\ref{WB-discr}) shows that the effective
bandwidth is $\Delta =1/T_{0}$ in the digital spectral analysis
with the Barttlet window.

Comparing Eqs.(\ref{Dig-base})-(\ref{W-FT}) with
Eqs.(\ref{Eq-base}) and (\ref {ILF-analog}) we can easily verify
that in the case of digital processing of
ILF noise measurements, an erroneous observation of the spectrum of the $%
1/f^{n}$ type will also take place. The digital measurements are
always carried out with a constant bandwidth ($\sim1/T_{0}$), at
least in each of several frequency intervals into which the entire
investigated band is subdivided. As a result of effective
''averaging'' of the resultant curve over all the sub-bands, a
spectrum of the type $1/f^{n}$ can be obtained with $1<n<2$ for
Barttlet window under conditions of ILF-noise measurements.

\subsection{An illustrative example}

\begin{figure}
  \includegraphics[height=.3\textheight]{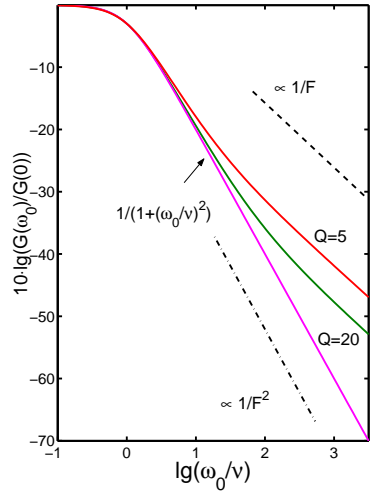}
  \caption{Dependence of the
spectrum analyzer output on the tuning frequency, Eq.  (11), under
condition of the constant quality factor $Q=f_{0}/\Delta $. Two
typical values of $Q$ are presented corresponding to the
equivalent noise bandwidth of 20\% and 5\%, respectively. The
spectral density of the noise at the analyzer input, Eq. (10), is
also shown for comparison.}\label{Fig}
\end{figure}

The observable signature of ILF noise measurements is shown in the
figure \ref {Fig}.
It was calculated in the case of Lorentzian form of both the noise
spectral density and the narrow-band input filter of the spectrum
analyzer.
The following expressions were substituted in Eq. (\ref{Eq-base}) as the $%
{\cal K}(\omega )$, $\Delta $ and $G_{y}(\omega )$%
\begin{eqnarray}
{\cal K}(\omega _{1}-\omega ) &=&\frac{\gamma ^{2}/4}{(\omega
_{1}-\omega )^{2}+\gamma ^{2}/4},\,\,\omega _{1}^{2}=\omega
_{0}^{2}-\gamma
^{2}/4,\,\Delta =\gamma /4  \label{K-Lorentz} \\
G_{y}(\omega ) &=&\left\langle y^{2}\right\rangle \frac{2\nu
}{\omega ^{2}+\nu ^{2}}  \label{Gy-Lorentz}
\end{eqnarray}
to obtain the explicit formula for the spectrum analyzer output
\begin{equation}
G^{out}(\omega _{1})=\left\langle y^{2}\right\rangle \frac{2\left(
\nu +\gamma /2\right) }{\omega _{1}^{2}+\left( \nu +\gamma
/2\right) ^{2}}. \label{Gout-Lorentz}
\end{equation}
The parameters $\omega _{0}$ and $\gamma $ are defined by the
equation of
the simple resonant filter\footnote{%
To compare Eqs. (11) and (12) with corresponding Eqs. (10) and
(12) in [2] one has to correct the misprint there and change
$\gamma \rightarrow \gamma /2$ taking into account the condition
(4).}
\[
\ddot{x}+\gamma \dot{x}+\omega _{0}^{2}x=y(t).
\]

It is easily seen that the condition (\ref{Q-defin}) for the
analyzer to be narrow-band (in our case it takes the form $\pi
\gamma /2\omega _{0}\ll 1$) still does not determine uniquely the
$G^{out}(\omega _{1})$ dependence. It is also important to know
the relation between $\gamma $ and $\nu $, i.e., between the
widths of the input spectrum and the resonant filter. At $\gamma
/\nu \ll 1$, i.e., in the case of a relatively narrower filter, we
have the usual result $G^{out}(\omega _{1})=G_{y}(\omega _{1}).$
In the other limiting case, however, when $\gamma /\nu \gg 1$, we
obtain
\begin{equation}
G^{out}(\omega _{1})=\left\langle y^{2}\right\rangle \frac{\gamma
}{\omega _{1}^{2}}\equiv \left\langle y^{2}\right\rangle
\frac{4\Delta }{\omega _{1}^{2}}  \label{ILF-Lorentz}
\end{equation}

From Eq. (\ref{ILF-Lorentz}) it follows directly that if the
measurements are carried out in this situation at a constant $Q$,
i.e., under the condition $\gamma /\omega _{1}=$ $const(\omega
_{1})$, then the output signal will imitate the presence of noise
with a spectrum $1/f$ at the input of the analyzer. It is clearly
seen in Fig.\ref{Fig} that although at high frequencies ($\omega
_{1}\gg \nu $) the true spectral density of the noise
(\ref{Gy-Lorentz}) is everywhere proportional to $1/\omega ^{2}$
and the condition (\ref{Q-defin})
is satisfied, all this is still not enough to let the region of the maximum $%
{\cal K}(\omega _{1}-\omega )$ to make the main contribution to
the integral in Eq. (\ref{Eq-base}). Conversely, as soon as the
filter bandwidth $\gamma $\ becomes larger than the width of the
input spectrum $\nu $, a transition from the expression
(\ref{Gy-Lorentz}) to (\ref{ILF-Lorentz}) takes place quite
rapidly. In other words, the output signal of the analyzer ceases
to yield information on the form of the spectrum of the input and
begins to vary with frequency in accordance with the $1/f$ law.

The measured curves of such a kind can be found in literature
without a discussion of possible ILF-noise interpretation. As a
result, some speculations take place in order to explain the
observed transition of measured spectral density from $1/f^{2}$ to
$1/f$ either by means of an unusual frequency dependence of the
sample properties (see, for instance \cite{Kim-Ziel-80}) or a
hypothesis is offered based on a decomposition of measured noise
spectrum on the sum of Lorentzian and $1/f$ components originated
by different causes \cite{L+1/f--2003}.

\section{$1/f$ noise in metal films}

Investigations of the excess current noise in metal films purposed
to shift the study of unclear $1/f$ noise onto the objects that
should be physically more defined than non-uniform semiconductors
or thermionic emission. If refer a start of that researches into
the beginning of 1970th, then it is to be ascertained that more 30
elapsed years and numerous new obtained data have not led to the
decision of a problem of origin of such a spectrum. Although
observed dependencies of spectral density of resistance
fluctuations at any fixed frequency on various parameters of a
sample material or measurement conditions point out as if these
variables are related with $1/f$ noise, it is failed to deduce
from here such its statistical properties as the spectral form or
the behavior of spatial correlations. The approach from the point
of view of ILF-noise allows to remove an acuteness of some
contradictions and restate the problem for additional experimental
studies.

\subsection{The temperature fluctuations}

Temperature fluctuations as the cause of $1/f$-noise have been
rejected by three principal arguments \cite{ShK
UFN85p285}-\cite{Weiss RMP88p537}: 1. The impossibility to deduce
the $1/f$ spectral density for fluctuations of the volume-averaged
film temperature. 2. The absence of correlations of current
fluctuations in spatially dividual parts of the film. 3. An effect
of the substrate on the measured resistance fluctuations. Let us
consider these conclusions from ILF-noise point of view.

To analyze the Voss and Clarke experiments it is necessary to
address to Eqs.(\ref{Dig-base}) and (\ref{WB-discr}). Assuming
$G_{y}(\omega )$ is ILF-noise and $\omega $ is not too close to
Nyquist frequency $\omega _{N}$ we have for the relative voltage
fluctuations \cite{AShJETP81}
\begin{equation}
\frac{G_{U}^{+out}(f)}{U^{2}}=\frac{2\sin ^{2}\left( N\pi \tau
_{0}f\right) }{\pi ^{2}\left( f/\Delta \right) }\frac{\beta
^{2}\left\langle \delta
T^{2}\right\rangle }{f}=\frac{1}{\pi ^{2}\left( f/\Delta \right) }\frac{%
\beta ^{2}T^{2}}{3N_{a}f}.  \label{ILF-Voss-Clarke}
\end{equation}

Here $G^{+}(f)=2G(f)$ is the spectral density in $f>0$ frequency semi-axis, $%
\beta =d\ln R/dT$, $R$ is the resistance of the film,
$\left\langle \delta T^{2}\right\rangle =k_{B}T^{2}/C$ is the mean
square of the equilibrium temperature fluctuations averaged over
the film volume, $C$ is the film heat capacity. In the case of the
room temperatures and monatomic metals $C=3k_{B}N_{a}$ and $N_{a}$
is the number of atoms in the film. The substitution of
experimental data \cite{V-C PRB76p556} combined with the hand-book
material parameters for the Bi and Au films in Eq.
(\ref{ILF-Voss-Clarke}) and hand-made estimations of
$\Delta_{Bi}=2\,Hz$ and $\Delta _{Au}=1\,Hz$ results
in\cite{AShJETP81}
\[
G_{Bi}^{+out}(10\,Hz)=15\cdot
10^{-17}Hz^{-1},\,\,G_{Au}^{+out}(10\,Hz)=0.58\cdot
10^{-17}Hz^{-1}
\]

The measured values are $13\cdot 10^{-16}Hz^{-1}$ (Bi) and
$0.6\cdot 10^{-16}Hz^{-1}$ (Au). The difference just in 10 times
for both cases suggests that there might be some problem with
conversion of digital Fourier-transform coefficients in $1/Hz$
spectral density. Anyway, the obtained of-order-of-magnitude
coincidence is worthy of notice, taking into account the lack of
fitting parameters and indeterminacy of the film dimensions
mentioned in \cite{V-C PRB76p556}. The $1/f^{\alpha }$ dependence
of the analyzer output with $1<\alpha <2$ follows from Eq. (\ref
{ILF-Voss-Clarke}) as it is explained in Sec.2.2.

As for the lack of spatial correlations of the volume-averaged
variables, it is necessary to note, first of all, that the
dependence of the relative correlation function like in Eq.
(\ref{ILF-Voss-Clarke}) on the sample volume $\Omega $ in the form
of $1/\Omega $ may occur only if the characteristic scale $r_{c}$
of the spatial correlation function $\left\langle \delta
T(r,t)\delta T(r_{1},t_{1})\right\rangle $ is much less than the
dimensions of the sample. The discussion of this point can be
found in \cite {Lif-Pit2001} from general point of view and in
\cite{ShK UFN85p285} with reference to $1/f$-noise problems.
Therefore, the spectral density of the frequency Fourier
components should be spatially correlated if they are formed from
the spatial Fourier components with characteristic wavelength of
order of the film dimensions. Thus there is no sense to look for
the spatial correlations between different parts of the film if
the $1/\Omega $ dependence of correlation function is observed and
vice versa.

This receipt regards to the case of the usual spectral
measurements. In the case of ILF-noise the output signal is
proportional to the mean square of temperature fluctuations taken
in the same instant time. In an equilibrium macroscopic body out
of phase transition region such fluctuations are always
uncorrelated on macroscopic scales in accordance with the
principle of additivity of extensive variables in thermodynamics.
The explicit prove of this statement for the spatial-averaged
temperature fluctuations can be found in \cite{LVain JETP82p1841}.
So in the case of ILF-noise the absence of spatial correlations
can not be an argument against of the temperature-fluctuation
origin of $1/f$ -noise.

To consider the dependence of $1/f$ noise on the substrate
supporting the film and on a degree of the film-substrate thermal
contact one need to account for the last stage of the time
relaxation of the averaged temperature fluctuations, i.e. a heat
interchange with the environment. The Newton boundary condition
for the thermoconductivity equation gives rise to the temperature
relaxation with one characteristic time $\tau $, resulting in the
simple expression for the spectral density of the form
\begin{equation}
G_{T}(\omega )=\left\langle \delta T^{2}\right\rangle \frac{2\tau
}{1+\omega ^{2}\tau ^{2}}  \label{G-T Lorentz}
\end{equation}

As $\tau $ is very large in comparison with the reciprocal lowest
measured frequency the ILF noise should take place and the
dependence on the substrate in Eq. (\ref{ILF-Voss-Clarke}) comes
from the heat capacitance $C$, which is determined by all
structure under study including an involved part of the substrate.
The latter depends also on the heat-transfer factor of the
film-substrate interface and can increase as $\tau $ decreases,
giving rise to the decrease of $G^{out}(\omega )$. As $1/\tau $
gets into the measured frequency range the usual spectral
measurements takes effect and $1/f$ behavior of the spectral
density can give place to an ordinary one.

So both characteristic dependencies of spectral measurements on
the substrate, which were discussed in \cite{D-H RMP81p497}
without definite conclusions, can be understood on the base of
ILF-noise model within the scope of the temperature fluctuations'
theory as well as two other problems considered in this section.
One can believe it is premature to discard the temperature
fluctuations as the source of $1/f$ noise in metal films.

\subsection{The defect-motion model}

After Eberhard and Horn have found out the activation dependence
of the $1/f$ -noise on the metal film temperature \cite{Eb-Ho
PRB78p6681} the opportunity of $1/f$-noise generation by defects
of the lattice was taken into consideration. As usual in the field
of flicker noise there is a contradiction between the observable
well-defined activation energy and the necessity to have a
wide-spread range of these energies to produce $1/f$ spectral
dependence over an extended frequency range. It was note in \cite
{Eb-Ho PRB78p6681} there is no possibility to avoid this
contradiction using only one activation process. It was suggested
to divide the defect contribution into the current noise by two
parts connected with the fluctuations of the defect number due to
a vacancy creation/annihilation and the fluctuation of the
free-carrier-scattering cross-section due to a defect motion with
different activation energy. The last process should be giving the
bread distribution of the activation energy. However, the observed
noise activation energy in various metal films are less than known
vacancy formation energy approximately to 10 times \cite{Eb-Ho
PRB78p6681}. Hence the observed activation energies should be
related with the motion of defects, leaving no place for the usual
construction of $1/f$-noise model.

Kogan and Nagaev \cite{ShKog-Nag SST84p387} suggested to connect
the motion-defect contribution to the flicker noise with some
characteristics of the internal friction process. The reasonable
estimation of possible noise level was obtained for that mechanism
(see also \cite{Weiss RMP88p537}) but the origin of the wide
distribution of the activation energy remained unexplained
\cite{ShK UFN85p285}. The ILF-noise model does not required any
real $1/f$ spectrum of fluctuations. It should be sufficiently
low-frequency only. The internal friction processes rather meet
this requirement having
characteristic relaxation frequencies in the diapasons of $%
10^{-12},10^{-8},10^{-5}$ Hz with magnitudes of reciprocal quality
$Q_{\max }^{-1}$ in the $10^{-1}\div 10^{-2}$ range \cite{Fink
Encyclo88}. One has to recall that the value of
$Q^{-1}\thickapprox 10^{-4}$ for the background level of the
internal friction was used in \cite{ShKog-Nag SST84p387} with a
positive conclusion.

\section{Concluding remarks}

It is worthy to mention the work \cite{Tinch PL85p357} where the
magnitude of $1/f$ noise in SQUIDs was evaluated as a result of
ILF-fluctuations of the current in superconducting or normal metal
shield surrounding the SQUID. The results obtained there keep
their significance for high-temperature SQUIDs, too
\cite{Ko-Clarke RMP99p631}.

It seems evidently that the ILF-noise model could remove
contradictions in existing theories of $1/f$ noise and should be
involved in the analysis of experimental results in the field of
flicker-noise investigations. A necessary change in the
experimental technique has been shortly discussed in
\cite{AShTPL81}.
\begin{theacknowledgments}
This work was partially supported by Russian Foundation for Basic
Researches.
\end{theacknowledgments}

\end{document}